

\long\def\UN#1{$\underline{{\vphantom{\hbox{#1}}}\smash{\hbox{#1}}}$}
\def\D{\partial}
\def\NL{\hfill\break}
\def\NI{\noindent}
\magnification=\magstep 1
\overfullrule=0pt
\hfuzz=16pt
\voffset=0.1 true in
\vsize=9.2 true in
    \def\NP{\hfill\break}
    \baselineskip 12pt
    \parskip 2pt
    \hoffset=-0.2 true in
    \hsize=6.6 true in
\nopagenumbers
\pageno=1
\footline={\hfil -- {\folio} -- \hfil}
\headline={\ifnum\pageno=1 \hfill Journal of Statistical Physics (1992) \fi}

\hphantom{AA}

\hphantom{AA}

\hphantom{AA}

\centerline {\bf MODEL OF CLUSTER GROWTH AND PHASE SEPARATION:}

\centerline {\bf EXACT RESULTS IN ONE DIMENSION}

\vskip 0.4in

\centerline{\bf Vladimir Privman$^{a,b,c}$}

\vskip 0.2in

\NI{$^a${\sl{Department of Physics, Theoretical Physics, University
of Oxford}}}\hfill\break{\hphantom{$^b$}{\sl{1, Keble
Road, Oxford OX1 3NP, UK}}}

\NI{$^b$on leave of absence from {\sl{Department of Physics,
Clarkson University}}}\hfill\break{\hphantom{$^c$on leave of
absence from} {\sl{Potsdam, NY 13699--5820, USA}}}

\NI{$^c$electronic mail: {\sl privman@craft.camp.clarkson.edu}}

\vskip 0.4in

\NI {\bf KEY WORDS\ }$\;$ Interface Diffusion, Phase Coexistence
and Growth, \NL \hphantom{\bf KEY WORDS }$\;$
Cluster Coarsening, Structure Factor Scaling

\NP

\centerline{\bf ABSTRACT}

We present exact results for a lattice model of cluster growth in $1D$.
The growth mechanism involves interface hopping
and pairwise annihilation supplemented by spontaneous creation
of the stable-phase, +1, regions by overturning the unstable-phase,
$-1$, spins with probability $p$.  For cluster coarsening at
phase coexistence, $p=0$, the conventional
structure-factor scaling applies. In this limit our model falls in the
class of diffusion-limited reactions A$+$A$\to$inert. The
$+1$ cluster size grows diffusively, $\sim \sqrt{t}$, and the
two-point correlation function obeys scaling. However, for $p>0$,
i.e., for the dynamics of formation of stable
phase from unstable phase, we find that structure-factor
scaling breaks down; the length scale associated with the size
of the growing
$+1$ clusters reflects only the short-distance properties of the
two-point correlations.

\NP

\NI{\bf 1. INTRODUCTION}

Lattice cellular automaton-type models with local tendency for ordering,
termed voter models, can be used to study phase segregation and
cluster coarsening reminiscent of spinodal decomposition,$^{(1,2)}$
at least in low dimensions. Both the cluster-size$^{(3)}$ and
structure-factor$^{(4)}$ scaling at phase separation have been subjects
of numerous investigations. However, most of the available results for
realistic $2D$ and $3D$ dynamical models are
numerical. We distinguish between the two ``scaling''
terms as follows. By
cluster-size scaling we mean scaling properties of the cluster size
distribution. The term structure-factor scaling is
reserved for the scaling
properties of the two-point order parameter
correlation function. The latter is
accessible to scattering experiments.

The symmetric voter-type models are related also to the
diffusion-limited chemical reactions involving particle annihilation,
A$+$A$\to$inert.$^{(5)}$
There are several exact results available mostly in $1D$ which
essentially translate to various average and asymptotic properties
of the cluster size distribution in the phase-separation
nomenclature.$^{(5-14)}$ Recent works also yielded exact results for
the two-point correlations.$^{(13,14)}$  Results for the chemical
reaction models in $D>1$ are more limited: see Ref.~10 and literature
cited therein. Furthermore, the relation of chemical reaction
systems to voter models is less straightforward.$^{(1,15-17)}$

The purpose of the present work is to introduce a lattice
model that incorporates voter-type cluster coarsening
by interface diffusion in $1D$, as well as the process of
spontaneous formation of stable-phase regions from those of the
unstable phase. We derive exact results for the two-point correlations.
Our main finding is that structure-factor scaling ideas cannot be
extended from cluster coarsening (of both phases)
\UN{at coexistence} to stable-phase
cluster growth \UN{off coexistence}. While the general-$D$ formulation
is outlined, the present study is focused on the $1D$ case.

The model is defined in Section~2. The generating function solution
of the discrete-time and discrete-space dynamics is presented in
Section~3. Section~4 is devoted to the discussion of some special
limits including the symmetric case. Our results are consistent with
previous studies; the general framework of our formulation is close
to the zero-temperature kinetic Ising model studies of $1D$
chemical reactions.$^{(8,13,14)}$ Detailed results are obtained
in the appropriately defined continuum limit of the discrete
dynamics (Section~5). These results are analyzed (Section 6)
with emphasis on the
length scales associated with the two-point correlation function.
The structure-factor scaling at coexistence and its breakdown
off coexistence are elucidated.

\NP

\NI{\bf 2. DEFINITION OF THE MODEL}

In this section we define the model in $1D$. We
also describe the extension to $D>1$. However, the emphasis in this
work is on the $1D$ case, and the notation is
introduced correspondingly.
Thus, we consider spin variables $\sigma_i (t) = \pm 1$. Time
evolves in unit steps:  $t=0,1,2,\ldots$. It is convenient to put
the spin variables only at even lattice sites, $i=0, \pm 2, \pm 4,
\ldots$ for even times $t=0,2,\ldots$. Similarly, for odd times $t=1,3,
\ldots$ we put spins at odd lattice sites $i=\pm 1, \pm 3, \ldots$.

The values $\sigma_i (t+1)$ for $(t+1)>0$ will be determined
stochastically by the dynamical rule incorporating interface diffusion
and pairwise annihilation
leading to cluster coarsening, and also spontaneous formation
of the stable, $+$, phase from the unstable, $-$, phase thus attempting
to model cluster growth in nucleation processes.
If the ``parent'' spins,
$\sigma_{i-1} (t)$ and $\sigma_{i+1} (t)$ are both $+1$ or both $-1$,
then the ``offspring'' is first set to $+1$ or $-1$, respectively.
However, the $-1$ value is then overturned with probability $p$. If the
``parent'' spin values are opposite, the ``offspring'' is first set to
one of them randomly. However, the $-1$ value is
again overturned spontaneously with probability $p$.

The first updating step (corresponding to setting $p=0$) describes a
symmetric ``voter model'' type dynamics.$^{(1,2,6,12)}$
Indeed, the ordered, all $+$ or all $-$, regions are unaltered.
However, each interface between the neighboring
$+$ and $-$ spin pairs hops one lattice spacing
to the left or to the right with
equal probability. On encounter, interfaces annihilate pairwise leading
to cluster coarsening. Note that interfaces can be viewed as located
at the odd sublattice at even times and at the even sublattice
at odd times. An important property of symmetric
voter models and related particle (here, interface)
diffusion-with-annihilation models is the
decoupling of the hierarchy of recursion relations for the correlation
functions,$^{(1,6)}$ as well as the diffusive nature of the
resulting equations in the continuum limit,
which have allowed derivation of several exact results.$^{(5-14)}$

The second step of updating, i.e., the spontaneous spin-flips $-1 \to
+1$, is introduced here as the means to break the $\pm$ symmetry and
model formation of the $+$ phase by growth
from the $-$ phase. Thus, we will
be interested in the results for $p \ll 1$. The key observation
(not limited to $D=1$) is that the correlation function hierarchy
can be set up in such a way that decoupling properties reminiscent
of the symmetric case are obtained.

At each lattice site and for each time $t>0$
we introduce two random variables, $\zeta_i (t)$
that takes on values 0 or 1 with equal probability, and $\theta_i (t)$
which is 0 with probability $(1-p)$ and 1 with probability $p$.
The stochastic dynamics is defined by

$$ \sigma_i (t+1) = \left[ 1 - \theta_i (t+1) \right]
\left\{ \zeta_i (t+1) \sigma_{i-1} (t) +
\left[ 1 - \zeta_i (t+1) \right] \sigma_{i+1} (t) \right\}
\, + \, \theta_i (t+1) {\rm \hphantom{AAAA}}
\eqno(2.1) $$

\NI Given that all the random variables $\zeta$ and $
\theta$ are statistically independent, one can easily verify
that the rule (2.1) correctly incorporates the dynamics as described
in the preceding paragraphs.

In calculating the averages, we can use the properties
$\sigma^2=1$, $\zeta^2=\zeta$, $\theta^2=\theta $ at fixed
time and lattice
coordinate. Furthermore, $\overline{\zeta_i (t)} = {1 \over 2} $, \ $
\overline{\theta_i (t)} = p$, where
the overbars denote statistical averages.

However, we still have to specify the initial values at $t=0$.
Either the values $\sigma_i (0)$ can be given deterministically
or quantities
involving the later-time values $\sigma_i (t>0)$
can be averaged over the distribution of the
initial conditions. Here we prefer the latter option; we assume that
the initial values are random and uncorrelated, with
$\overline{\sigma_i (0)} = \mu$. Thus, $-1\leq \mu \leq 1$ is
the initial magnetization: each
$\sigma_i (0)$ takes on values $+1$ and
$-1$ with respective probabilities $(1+\mu )/2$ and $(1-\mu )/2$.

Let us now define the average quantities that will be considered
in this study. Firstly, due to translational invariance of the initial
conditions and of the dynamical rule (2.1) after averaging over the
random variables, the magnetization, $m(t)=\overline{\sigma_i (t)}$,
depends only on time and satisfies the recursion ($t \geq 0$)

$$ m(t+1)=(1-p)m(t)+p \eqno(2.2)$$

\NI with

$$ m(0)=\mu \eqno(2.3) $$

\NI Similarly, the two-point correlation function,

$$ G_n (t) = \overline{\sigma_i (t) \sigma_{i+n} (t)} \eqno(2.4) $$

\NI depends only on the distance $n=0,2,4,\ldots$ between the spins.
The recursion relation for the two-point function for $n>0$ and $t>0$
is easily derived from (2.1):

$$ G_n(t+1)={1\over 4}(1-p)^2 \left[ G_{n-2} (t) + 2G_n (t)
+ G_{n+2} (t) \right] +2p(1-p)m(t)+p^2 \eqno(2.5) $$

\NI where the initial and boundary conditions are

$$ G_{n=0} (t\geq 0)=1 \;\;\; {\rm and}
\;\;\; G_{n>0}(t=0)=\mu^2 \eqno(2.6) $$

The decoupling of the equations for the correlation functions
should be obvious at this stage. Due to the linearity of (2.1) in
$\sigma$, the $k$-point averages at $(t+1)$ are determined by the $k$,
$k-1$, $\ldots$-point averages at time $t$. This property is further
amplified for
the \UN{connected} correlation function defined by

$$ C_n (t) = G_n (t) - m^2(t) \eqno(2.7) $$

\NI Indeed, the appropriate recursion is

$$ C_n(t+1)={1\over 4}(1-p)^2 \left[ C_{n-2} (t) + 2C_n (t)
+ C_{n+2} (t) \right] \eqno(2.8) $$

\NI so that the $m(t)$-dependence enters
only via the boundary conditions. The equivalents of relations (2.6) are

$$ C_{n=0} (t\geq 0)=1-m^2(t) \;\;\;
{\rm and} \;\;\; C_{n>0}(t=0)=0 \eqno(2.9) $$

\NI where we used (2.3).

The probability to find an interface at $i$, in the interstice
between the two spins $\sigma_{i\pm 1}$, is given by

$$ \rho (t) = {1 \over 2}\left[1-G_2 (t)\right]
={1\over 2}\left[ 1-m^2(t)-C_2(t) \right]
={1\over 2}\left[ C_0(t)-C_2(t) \right] \eqno(2.10)$$

\NI Similarly to $m(t)$, this quantity is translationally invariant.
Both $m$ and $\rho$ can be also considered as the order-parameter and
interface-number
\UN{densities} if we allow for the fact that they  are
defined per site of the lattice of twice the spacing of the original
$1D$ linear system of sites labeled by $i$. In fact, all our
calculations will be with dimensionless variables such as distance $n$
and time $t$. One can of course introduce dimensional length and time
scales which has been a common practice especially in the continuum
limit. However, we found that no new useful physical insight in gained,
while the equations become more complicated. Thus, we use the
dimensionless variables throughout.

Before we outline the extension to $D>1$,
which will be detailed elsewhere, let us emphasize
three appealing features of the $1D$ model: the
property that only two parent spins ``vote'' at each
time step, the linearity of the evolution rule (2.1), and the fact
that for $p=0$ there are already many results available, in particular,
the relation to the interface motion and the interpretation of cluster
coarsening due to interface annihilation.

The simplicity of two-spin voting and the linearity of the
dynamical rule (implying, essentially, solvability)
can be extended to $D>1$ by using the
idea of updating along different axes in each time step.$^{(2)}$
Consider spins $\sigma_{i_1 i_2 \ldots i_D} (t)$.
For time steps $t = 0 \to 1$, $D \to D+1$, $2D \to 2D+1$, $\ldots$,
the rule (2.1) is used along axis 1, i.e., with $i_1$ varied as in
(2.1) while $i_2,\ldots ,i_D$ kept the same on both sides of the
relation. Similarly,
for time steps $ 1 \to 2$, $D+1 \to D+2$,  $\ldots$,
the update relation involves the index $i_2$ along axis 2, and so on.
In $D$ time steps, the cycle of the axis indices is complete.

Regarding the availability of exact results for
$p=0$ and the interpretation of the dynamics
of the broken bonds connecting $\pm$ spin pairs,
the $D>1$ results, see Ref.~10 and literature cited therein,
are understandably less numerous than those available for $D=1$.
In fact, it has been argued$^{(1,15-17)}$ that symmetric
voter-model type dynamics cannot lead to cluster coarsening
in $D=3$ and higher. In $D=2$ the clusters do grow$^{(1,15-17)}$
but the process can no longer be
described by a simple cluster-size scaling.$^{(1,2)}$
Quite generally, many open questions remain for $D>1$.

\NP

\NI{\bf 3. GENERATING FUNCTION FORMULATION}

The recursion relation (2.2) for the magnetization is trivial to solve,

$$ m(t)=1-(1-\mu )(1-p)^t \eqno(3.1) $$

\NI However, the solution for the correlation function can be obtained
in a simple form only in terms of the generating functions

$$ B_n(v)=\sum_{t=0}^\infty v^t C_n(t) \eqno(3.2) $$

\NI Indeed, relations (2.8) yield, for $n=2,4,\ldots $,

$$ B_n={v \over 4}(1-p)^2 \left(B_{n-2}+2B_n+B_{n+2} \right)
\eqno(3.3) $$

\NI while for $n=0$ we get

$$ B_0(v)={2(1-\mu ) \over 1-(1-p)v}-{(1-\mu )^2 \over 1-(1-p)^2v}
\eqno(3.4) $$

\NI where we used the conditions (2.9) in deriving both (3.3)
and (3.4).

The second-order difference equation (3.3) has two linearly
independent solutions of the form

$$ B_n(v) \propto b^{n/2} \eqno(3.5) $$

\NI where $b(v)$ is a root of the quadratic characteristic
equation. However, one can check that only one of the two roots
yields the solution which converges exponentially to zero as
$n \to \infty$ (for fixed $v$ is the vicinity of 0). The other root
yields exponentially divergent terms. After some algebra we arrive
at the expression

$$ B_n(v)=B_0(v) \left\{ { \left[1-
\sqrt{1-(1-p)^2 v} \right]^2 \over (1-p)^2 v} \right\}^{n/2}
\eqno(3.6) $$

\NI where $n=2,4,\ldots$.

The result (3.6), when expanded in powers of $v$,
yields $C_n(t)$ as the $t^{\rm th}$ Taylor
series coefficient. However, the expressions thus obtained involve
double sums and are rather unilluminating. The continuum limit
results derived in Sections 5 and 6 provide a more useful source of
physical insight on the nature of the dynamics.

Our main interest presently will be in the expression of
the generating function for the interface density
$\rho(t)$, see (2.10). This quantity is the $t^{\rm th}$
Taylor coefficient of the function

$$ {1 \over 2} \left[ B_0(v)-B_2(v) \right] \eqno(3.7) $$

\NI Explicit calculation yields the result

$$ \rho (t)=  2(1-\mu )(1-p)^t \left[1- S_t(1-p)\right]
-(1-\mu )^2 (1-p)^{2t} \left[1- S_t(1)\right] \eqno(3.8) $$

\NI where we defined the finite sum

$$S_t (\alpha) = \sum_{k=0}^t \, {(2k)! \, \alpha^k
\over k! \, (k+1)! \, 2^{2k+1} } \eqno(3.9) $$

\NI Note that

$$ S_\infty (\alpha)= \left( 1-\sqrt{1-\alpha} \right) \big /
\alpha \eqno(3.10) $$

\NI where the $t=\infty$ Taylor series converges for all
$\alpha$ in $[0,1]$.

\NP

\NI{\bf 4. DIFFUSION AS OPPOSED TO SPIN-FLIP}

It is instructive to consider two models which represent the extremes
of diffusion only or no diffusion at all, as far as interfacial dynamics
is concerned. If we set $p=0$ in our model, the only processes are
those of interface hopping and pairwise annihilation. Thus, the
model falls in the class of the diffusion-limited chemical reactions
A$+$A$\to$inert, which, as well as related models,
were studied extensively$^{(5-14)}$ in $1D$.
The density of interfaces reduces to

$$ {\rho (t) \over \rho(0)}=\sum_{k=t+1}^\infty \,
{(2k)! \over k!\, (k+1)!\, 2^{2k} } \;\;\;\;\;\;\;\;\;\;\;\; (p=0) \eqno(4.1)
$$

\NI where generally,

$$ \rho (0)=(1-\mu^2)/2 \eqno(4.2) $$

The density of interfaces decreases monotonically and smoothly
for discrete time steps $t=0\to 1\to 2\to \ldots$. For large times,
we have

$$ \rho(t) = 2\rho(0) \big / \sqrt{\pi t} \;\;\;\;\;\;
\;\;\;\;\; (p=0) \eqno(4.3) $$

\NI where the $\sim\! t^{-1/2}$ law is consistent with
the previous exact calculations for these reactions.

The other extreme would be to have no diffusion at all. For this,
however, we have to modify our model. Thus, let us consider a model
of $\pm 1$ spins with the only dynamical process
consisting of spin-flips $\, -1 \to +1$ with probability $p$.
Since the spins are uncorrelated, the model is trivial to solve.
Indeed, the dynamical equation (2.1) is replaced by

$$ \sigma_i (t+1) = \left[ 1 - \theta_i (t+1) \right]
\sigma_i (t) + \theta_i (t+1) \eqno(4.4) $$

\NI where we now assume that the spins are located on the even
sublattice at all times.

The magnetization obeys the same equation
(2.2), which simply reflects the fact that in our more complicated
model diffusion of interfaces conserves the order parameter.
However, in the new, uncorrelated-spin model (USM), all the $k$-point
correlations factorize trivially, and as a result the connected
correlations vanish identically (for distinct $k$ coordinates).
Specifically, we get

$$ \rho^{\rm USM} (t)=(1-m^2)/2=(1-\mu )(1-p)^t-{1\over 2}
(1-\mu )^2 (1-p)^{2t} \eqno(4.5) $$

$$ C_n^{\rm USM} (t)=\delta_{n,0}\, [1-m^2(t)] \eqno(4.6) $$

The time-dependent length scale of interest in cluster growth is
the average size of the dominant, $+$, clusters.
More generally, one may consider the cluster size distribution, which
was not obtained analytically.
(For some asymptotic results in the diffusion-only model see Ref.~6.)
One measure of the cluster size is $\rho^{-1} (t)$. For the
diffusion-only model this cluster size measure grows according to $\sim
\sqrt{t}$ for large times. For the USM, it grows as $\sim (1-p)^{-t}$
(assuming $0<p<1$).

Note, however, that this quantity is related to
the \UN{short-distance} properties of the two-point correlations,
see (2.10). The moment or
decay-tail definitions of the ``correlation''
length scales (defined in Section 6) are typically used to probe
the fixed-time \UN{large-distance} behavior of the two-point
correlations in strongly fluctuating systems.
The various length scales are not necessarily related.
For the diffusive model (symmetric, phase coexistence case, $p=0$),
it turns out that all the length scales behave according to
$n \sim \sqrt{t}$ (Section 6).  The USM example is
instructive as the opposite extreme: the two-point correlations are
zero-range; see (4.6). However, the cluster size measure $\rho^{-1}$
is well defined and diverges as $t \to \infty$.
The length scale properties will be further explored in Section 6.

\NP

\NI{\bf 5. CONTINUUM LIMIT}

The continuum limit has been the standard framework for
writing phenomenological equations in cases which are not
exactly solvable, or where the precise microscopic dynamics is not
known or specified. If fact, the continuum limiting description
provides a useful guide for the identification
of the ``universality classes'' or at least general classes of models
with similar properties. A simple-minded continuum limiting procedure
would amount to the assertion that for $t\gg 1$ and $n\gg 1$ the
discrete variation can be replaced by smooth functional dependence on
$t$ and $n$. Formally, one then uses the expression

$$ f(t+\Delta t,n+\Delta n)=\left\{
\exp \left[ \lambda \Delta t {\D \over \D t} \right]
\exp \left[ \Lambda \Delta n {\D \over \D n} \right]
f(t,n) \right\}_{\lambda=1,\, \Lambda=1} \eqno(5.1) $$

\NI to expand in the derivatives, which are presumably small.
The order of the expansion is conveniently monitored by collecting
powers of $\lambda$ and $\Lambda$ before setting these variables to 1
in the final expressions.

If we apply this procedure to the equation (2.2) for $m(t)$, and keep
the leading $t$-derivative, we obtain the equation

$$ {dm \over dt}=-pm+p \eqno(5.2) $$

\NI with the solution

$$ m(t)=1-(1-\mu ){\sl e}^{-pt} \eqno(5.3) $$

\NI where we used (2.3). However, this result differs from the exact
expression (3.1), which is, in fact, perfectly well defined for all
real $t\geq 0$. The source of the difficulty is clearly in that the
$t$-derivatives are small only for $p\ll 1$. This
is an illustration of the well-known property that the continuum
approximation can be used only in a limited part of parameter
space.

A better controlled procedure is to use properly rescaled variables
so that the parameters $\lambda$ and $\Lambda$ in the equivalent of
(5.1) are actually small. For our problem, we set

$$ \tau = p\, t \;\;\; {\rm and} \;\;\; \lambda=p \eqno(5.4) $$

$$ x = \sqrt{p}\, n \;\;\; {\rm and} \;\;\; \Lambda=\sqrt{p}
\eqno(5.5) $$

\NI where the $t$-rescaling is suggested by our consideration of $m(t)$,
while the $n$-rescaling is implied by the diffusive combination
$n/t^2$ which is expected to survive the $p \to 0$ limit.

In terms of the new variables, the relation (2.8) for the
two-point function has the leading terms in order $p$.
The ``continuum limit'' two-point function
$C(x,\tau)$ satisfies the relation obtained by collecting these
terms,

$$ {\D C \over \D \tau } = -2C + {\D^2 C \over \D x^2 }
\;\;\;\;\;\;\;\;\;\;\;\;\;\;\;\; (x>0) \eqno(5.6) $$

\NI where the initial and boundary conditions (2.9) are replaced by

$$ C(x=0,\tau \geq 0)=1-m^2(\tau )  \;\;\; {\rm and} \;\;\;
C(x>0,\tau =0) =0 \eqno(5.7) $$

\NI with

$$ m(\tau )=1-(1-\mu ){\sl e}^{-\tau } \eqno(5.8) $$

\NI The interface density in the continuum limit is approximated as
follows,

$$ {\rho \over \sqrt{p}} \simeq \left[- {\D C (x,\tau )
\over \D x } \right]_{x=0} \eqno(5.9) $$

The reader should keep in mind that the continuum limit is an
\UN{approximation} valid asymptotically
for $0\leq p \ll 1$, $\; t\gg 1$, $\; n\gg 1$.
The results must be properly interpreted. For instance, if taken
literally, relation (5.9) would imply that the interface density is
infinite at $\tau=0$ because the initial conditions for $C(x,\tau )$
are step-like. In fact, the divergence is in the regime
where the continuum limiting approximation breaks down; see the next
section. The rescaling (5.4)-(5.5) also obscures the $p=0$ case. Indeed,
the results must be properly expressed in terms of the variable
$x/ \tau^2 = n/t^2$ before taking the limit $p \to 0$. If fact,
$p=0$ is reminiscent of the ``critical-point'' limit in which there are
no small parameters to rescale $n$ and $t$. Instead,
only their ``scaling combination'' enters in the continuum limit.

The solution of the equation (5.6) with conditions (5.7) is obtained
by the Laplace Transform method. We omit the mathematical details
and only quote the final expression,

$$ \int\limits_0^\infty {\sl e}^{-\omega \tau } C(x, \tau) d \tau
=\left[ {2(1-\mu ) \over \omega +1} - {(1-\mu )^2 \over \omega +2}
\right] {\sl e}^{-\sqrt{\omega+2}\, x} \eqno(5.10)$$

\NI which inverse-transforms to

$$ {C(x, \tau ) \over 1-\mu } = {\sl e}^{-\tau } \left[
{\sl e}^x {\rm erfc}\! \left({x \over 2 \sqrt{\tau } } + \sqrt{\tau }
\right) +{\sl e}^{-x} {\rm erfc}\!
\left({x \over 2 \sqrt{\tau } } - \sqrt{\tau } \right)
-(1-\mu ) {\sl e}^{-\tau } {\rm erfc}\! \left({x \over 2 \sqrt{\tau } }
\right) \right] \eqno(5.11) $$

\NI where

$$ {\rm erfc}(\alpha )={2 \over \sqrt{\pi }
}\int\limits_\alpha^\infty {\sl
e}^{-\beta^2}d\beta \eqno(5.12) $$

\NI is one of the standard error functions,
the properties of which are well known. Thus, the expression (5.11) can
be used to analyze various
properties of connected two-point correlations. Some such results
will be presented in the next section.

\NP

\NI{\bf 6. LENGTH SCALES AND BREAKDOWN OF SCALING}

Let us consider the large-$x$ behavior of $C(x,\tau)$ for fixed
$\tau > 0$ (and $\mu \neq 1$). All three terms in (5.11)
then follow the asymptotic large-argument behavior of the error
function. The result turns out to be

$$ C(x\to \infty, \tau) \propto {\sl e}^{-2\tau } \left[
{\sqrt{\tau } \over x} \exp \left( - {x^2 \over 4\tau }\right)
\right] \eqno(6.1) $$

\NI where we omitted the proportionality constant.
The decay-tail length scale, $n_{\rm tail}$, is thus determined
by the dependence on the diffusional combination $x^2/\tau =n^2/t$,

$$ n_{\rm tail} \sim \sqrt{t} \eqno(6.2) $$

Consider next the moment-definition length scales. We define
the $k^{\rm th}$ moment,

$$ M_k(\tau )=\int\limits_0^\infty x^k C(x,\tau ) dx \eqno(6.3) $$

\NI and the associated time-dependent length, $n_k (t)$,

$$ \sqrt{p}\, n_k = \left( M_k \big / M_0 \right)^{1/k} \eqno(6.4) $$

\NI In the evaluation of $M_k$, the contribution due to the first
term in (5.11) can be always used in its large-argument form, while the
third term is originally a function of the diffusional combination
(times ${\sl e}^{-2 \tau }$). The second term, however, can be written
in such a diffusional-scaled form only for

$$ x \gg a_1 \sqrt{\tau} + a_2 \tau \eqno(6.5) $$

\NI where from now on the coefficient notation $a_j$ will be defined to
stand for ``a slow varying function of $\tau$, of order 1,
possibly $k$-dependent (when implied by context).'' The
diffusional contribution to the moments is

$$ M_k^{\rm (diff)} = a_3 \tau^{(k+1)/2}\, {\sl e}^{-2 \tau }
\eqno(6.6) $$

\NI In the range of smaller $x$, not satisfying (6.5), the error
function in the second term in (5.11) becomes of order 1. In fact,
the fixed-$x$, large-time behavior

$$ C(x, \tau\to\infty) \propto {\sl e}^{-\tau -x} \eqno(6.7) $$

\NI is explicit in the Laplace-transformed form (5.10)
due to the rightmost
pole singularity at $\omega =-1$. The added contribution due to this
exponential behavior is of the form

$$ M_k^{\rm (exp)} = a_4 {\sl e}^{-\tau } \int\limits_0^{
a_1 \sqrt{\tau} + a_2 \tau} x^k {\sl e}^{-x} dx \eqno(6.8) $$

\NI For small $\tau$, the intergation will yield the same power
of $\tau $ as in (6.6). Thus, the moment length scales
behave according to

$$ n_k \sim  \left(\tau^{(k+1)/2} \big / \tau^{1/2}\right)^{1/k}\,
p^{-1/2}= \sqrt{t} \;\;\;\;\;\;\;\; (t\ll 1/{p}) \eqno(6.9) $$

\NI However, as $\tau $ increases, the integral in (6.8) saturates
at a value
of order 1. Since the remaining time dependence, ${\sl e}^{-\tau }$,
dominates that of the diffusive contribution (6.6), the length scales
saturate at

$$ n_k \sim 1/\sqrt{p} \;\;\;\;\;\;\;\; (t\gg 1/{p}) \eqno(6.10) $$

\NI The crossover between the limiting behaviors occurs at $t \sim
1/p$ and is difficult to evaluate in closed form.

We next turn to the density of interfaces and the associated length
scale. A direct calculation of the right-hand side of (5.9) yields

$$ {\rho \over \sqrt{p}} \simeq
{1-\mu^2 \over \sqrt{\pi \tau } } {\sl e}^{-2\tau}
+2(1-\mu ){\sl e}^{-\tau} {\rm erf} (\sqrt{\tau})  \eqno(6.11) $$

\NI where we kept the approximation sign to emphasize that this result
applies only for $t\gg 1$ (as well as $p\ll 1$). Note that
${\rm erf} (\alpha)=1-{\rm erfc} (\alpha)$. The limit $p=0$ is thus
correctly reproduced; see (4.3). For small $\tau$,
the first term in (6.11) dominates, and the associated length scale
behaves according to

$$ n_\rho \propto \rho^{-1} \sim \sqrt{t}
\;\;\;\;\;\;\;\; (t\ll 1/p) \eqno(6.12) $$

\NI However, for large $\tau$ the second term takes over. Noting that
the function ${\rm erf} (\alpha )$ approaches
1 for large $\alpha$, we conclude that

$$ n_\rho \sim {\sl e}^{pt} \big / \sqrt{p}
\;\;\;\;\;\;\;\; (t\gg 1/p) \eqno(6.13) $$

In the theories of structure-factor scaling,$^{(4)}$
where the structure
factor is defined as the spatial Fourier transform of $C(n,t)$, assuming
continuous coordinate $n$ and time $t$, the momentum, $q$, dependence
is scaled in the form $\hat n (t) q$. In the direct-space notation
this amounts to assuming that the coordinate dependence enters
via $n/\hat n(t)$. It is tempting to associate $\hat
n(t)$ with a typical
cluster size measure. In practice, $\hat
n$ is determined as the inverse
of some momentum scale found at low or fixed $q$-values, corresponding
to large or intermediate coordinate values $n$.

Our results support this picture \UN{at coexistence}, i.e., at $p=0$.
Indeed, due to the critical-point-like scaling expressed by the
diffusive scaling combination $n^2/t$, all length scales defined at
short or large distances are essentially the same. The identification
$\hat n \sim \sqrt{t}$ is unambiguous.
However, explicit expressions obtained for $p > 0$
indicate two difficulties with the structure-factor
scaling when the growth of the stable phase occurs off coexistence.
Firstly, the identification of a unique length scale is no longer
possible for large times for which the cluster size distribution
deviates significantly from the symmetric case. All three length
scales estimated behave differently for large $t$. Secondly,
the two-point correlation function no longer has simple scaling
properties. In fact, a more general conclusion, alluded to in
Section~4, is that in such cases the length scale $n_\rho (t)$
is the appropriate one to use as a typical $+$ cluster size. However,
it is characteristic only of the \UN{short-distance}
coordinate dependence of the two-point function.

In summary, we presented a solvable $1D$ model of cluster growth.
Our results indicate that the ideas of structure-factor scaling
apply only to cluster coarsening at coexistence. Off coexistence,
a typical stable-phase cluster size measure
reflects only the short-distance
properties of the two-point correlations; the full correlation function
no longer obeys scaling.

The author acknowledges helpful
discussions with M.A.~Burschka, C.R.~Doering, D.A.~Rabson
and R.B.~Stinchcombe.  This research was partially supported by the
Science and Engineering Research Council (UK),
under grant number GR/G02741, and by a Guest Research Fellowship at
Oxford, awarded by the Royal Society.

\NP

\centerline{\bf REFERENCES}

\item{1.} M. Scheucher and H. Spohn, J.
Stat. Phys. {\bf 53}: 279 (1988).

\item{2.} B. Hede and V. Privman, J.
Stat. Phys. {\bf 65}: 379 (1991).

\item{3.} Review: P. Meakin, in {\sl Phase Transitions
and Critical Phenomena}, Vol. 12,
C. Domb and J.L. Lebowitz, eds. (Academic Press, New York, 1988),
p. 336.

\item{4.} Review: J.D. Gunton, M. San Miguel
and P.S. Sahni, in {\sl Phase
Transitions and Critical Phenomena}, Vol. 8, C. Domb and J.L.
Lebowitz, eds. (Academic Press, New York, 1983), p. 269.

\item{5.} Review: V. Kuzovkov and E. Kotomin, Rep. Prog. Phys.
{\bf 51}: 1479 (1988).

\item{6.} M. Bramson and D. Griffeath,
Ann. Prob. {\bf 8}: 183 (1980).

\item{7.} D.C. Torney and H.M. McConnell,
J. Phys. Chem. {\bf 87}: 1941 (1983).

\item{8.} Z. Racz, Phys. Rev. Lett. {\bf 55}: 1707 (1985).

\item{9.} A.A. Lushnikov, Phys. Lett. A{\bf 120}: 135 (1987).

\item{10.} M. Bramson and J.L. Lebowitz, Phys. Rev. Lett. {\bf 61}:
2397 (1988).

\item{11.} D.J. Balding and N.J.B. Green, Phys. Rev. A{\bf 40}:
4585 (1989).

\item{12.} J.W. Essam and D. Tanlakishani, in {\sl Disorder in Physical
Systems}, R.G. Grimmet and D.J.A. Welsh, eds. (Oxford University Press,
1990), p. 67.

\item{13.} J.G. Amar and F. Family, Phys. Rev. A{\bf 41}: 3258 (1990).

\item{14.} A.J. Bray, J. Phys. A{\bf 23}: L67 (1990).

\item{15.} P. Clifford and A. Sudbury, Biometrika {\bf 60}: 581 (1973).

\item{16.} R. Holley and T.M. Liggett, Ann. Prob. {\bf 3}: 643 (1975).

\item{17.} J.T. Cox and D. Griffeath, Ann. Prob. {\bf 14}: 347 (1986).

\bye